# Detecting Lifshitz Transitions Using Nonlinear Conductivity in Bilayer Graphene

Tanweer Ahmed,[1, *] Harsh Varshney,[2] Bao Q. Tu,[1] Kenji Watanabe,[3] Takashi Taniguchi,[4] Marco Gobbi,[5,6] Fèlix Casanova,[1,6] Amit Agarwal,[2, †] and Luis E. Hueso[1,6, ‡]

[1]CIC nanoGUNE BRTA, 20018 Donostia-San Sebastian, Basque Country, Spain.

[2]Department of Physics, Indian Institute of Technology, Kanpur 208016, India.

[3]Research Center for Electronic and Optical Materials, National Institute for Materials Science, 1-1 Namiki, Tsukuba 305-0044, Japan.

[4]Research Center for Materials Nanoarchitectonics, National Institute for Materials Science, 1-1 Namiki, Tsukuba 305-0044, Japan.

[5]Centro de F´ısica de Materiales (CFM-MPC) Centro Mixto CSIC-UPV/EHU, 20018 Donostia-San Sebastian, Basque Country, Spain.

[6]IKERBASQUE, Basque Foundation for Science, 48009 Bilbao, Basque Country, Spain.

The second-order nonlinear electrical response (NLER) is an intrinsic property of inversion symmetry-broken systems which can provide deep insights into the electronic band structures of atomically thin quantum materials. However, the impact of Fermi surface reconstructions, also known as Lifshitz transitions, on the NLER has remained elusive. We investigated NLER in bilayer graphene (BLG), where the low-energy bands undergo Lifshitz transitions. Here, NLER undergoes a sign change near the Lifshitz transitions even at elevated temperatures $T \geq 10K$. At the band edge, NLER in BLG is modulated by both extrinsic scattering and interfacial-strain-induced intrinsic Berry curvature dipole, both of which can be finely tuned externally by varying doping and interlayer potential. Away from the band edge, BLG exhibits second-order conductivity exceeding 30 $\mu mV^{-1}\Omega^{-1}$ at 3K, higher than any previous report. Our work establishes NLER as a reliable tool to probe Lifshitz transitions in quantum materials.

The influence of the Fermi surface topology [1] on the electronic properties of conductors can be observed through classical [2], quantum Hall [3] and anomalous Hall [4] effects under time-reversal symmetry broken conditions. In inversion symmetry-broken systems, scattering mechanisms [5–10] at the Fermi surface and Berry curvature [9,11–18] of the electronic bands contribute prominently to second-order nonlinear electronic responses (or NLER) [19,20] even without time-reversal symmetry breaking. Consequently, NLER can be sensitive to topological reconstructions [12,21] in the Fermi surfaces. Unlike traditional bulk conductors, the Fermi energy ( $E_F$ ) in atomically thin quantum materials can be precisely tuned across singularities in the electronic bands [22,23], known as Lifshitz transitions, where the topology of the Fermi surface changes. Recent advances suggest that NLER in two-dimensional (2D) van der Waals (vdW) crystals and their heterostructures can be exceptionally large, often surpassing that of traditional bulk crystals by orders of magnitude [6,7,24]. This enhancement opens new possibilities for NLER in logic circuits and energy-harvesting technologies [25,26]. However, there is a lack of both experimental and theoretical studies on NLER in systems where the band structure and Lifshitz transitions can be dynamically controlled by tuning thermodynamic parameters.

Bernal stacked bilayer graphene (BLG), whose low energy ($E$) Fermi surfaces undergo Lifshitz transitions as the $E_F$ is varied [27–29], is an ideal system for such studies. The inversion symmetry of free-standing BLG can be broken by hexagonal boron nitride (hBN) encapsulation and/or applying a vertical displacement field ($D$), which introduces a $D$-dependent band gap and Lifshitz transitions [27–29]. This setup fulfills the conditions necessary for scattering-mediated extrinsic NLER via side jump and skew scattering pathways [5–10,26,30]. Additionally, it leads to a finite Berry curvature [31,32], whose sign can be switched by altering the valley index ($K$ or $K'$), carrier type (electron (e) or hole (h)), and the sign of $D$. When combined with $C_3$ symmetry-breaking mechanisms such as interfacial strain, BLG can also exhibit a tunable Berry curvature dipole (BCD) [33], leading to intrinsic NLER [5,9,11–18,24]. Notably, the intrinsic NLER contributes only to the transverse component, while extrinsic contributions affect both longitudinal and transverse components. Previous studies have demonstrated large extrinsic and intrinsic NLER, in graphene based systems are sensitive to Fermi surface reconstructions [6,12,34]. However, the role of Lifshitz transitions in NLER, particularly from both extrinsic and intrinsic pathways in BLG, remains unexplored.

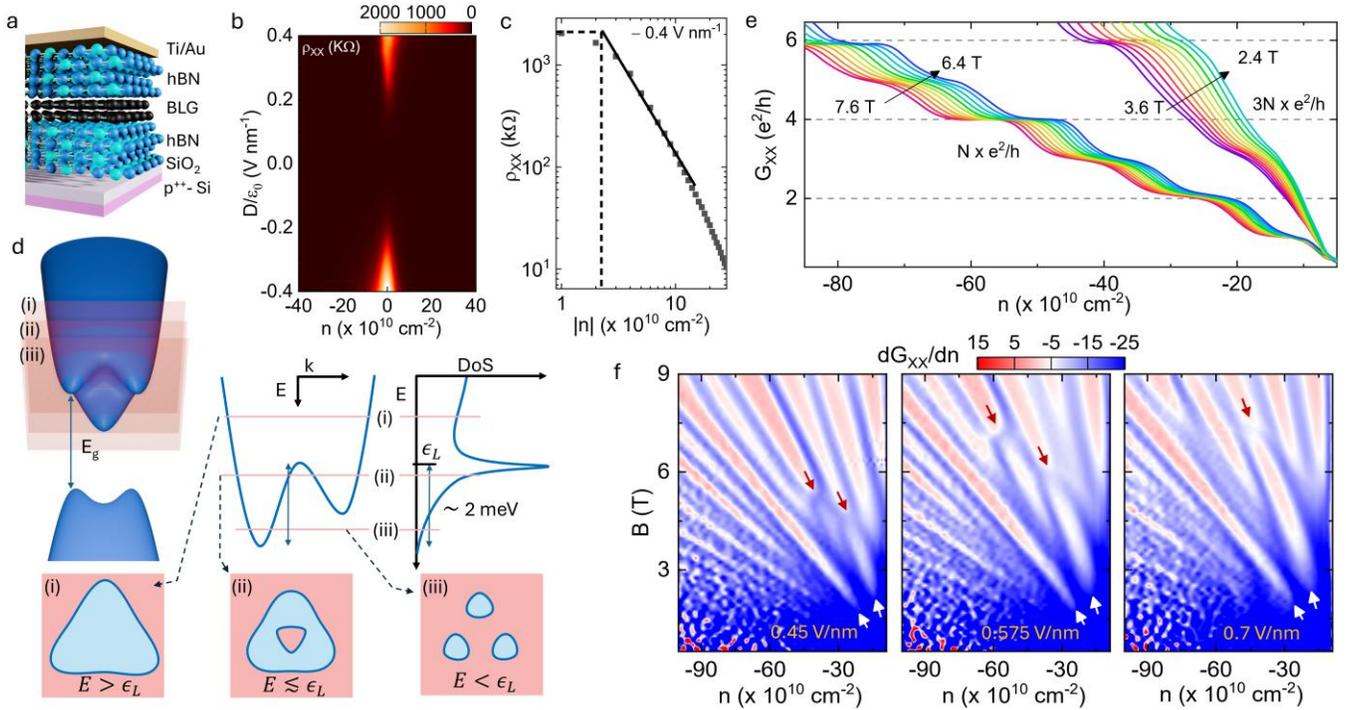

Figure 1. **Signature of Lifshitz transition in bilayer graphene from first order transport: a,** The scheme of the heterostructure is shown. **b,** $D - n$ phase space of resistivity. **c,** $\rho_{xx} - n$ data showing charge disorder density $\sim 10^{10}$ cm$^{-2}$ in gapped BLG at $D/\varepsilon_0 = -0.4$ Vnm$^{-1}$. **d,** Left panel schematically shows low energy conduction (CB) and valance (VB) of gapped BLG in the $E - k$ space. i, ii and iii indicate constant $E_F$ planes close to the band edge. The projections of the CB at constant $E_F$ planes are shown in the bottom, panels. This demonstrate the Lifshitz transition at $E = \epsilon_L$ (close to ii) The top middle panel show the vertical plane projection of the CB. The DoS as function of $E$ is shown in the top right panel. **e,** The conductivity ($G_{xx}$) vs $n$ at different values of vertical magnetic field ($B$). At larger $B$ (7.6 T to 6.4 T) quantized $G_{xx}$ plateaus at $N \times e^2/h$ values, where as at lower $B$ (3.6 T to 2.4 T), 3 fold degenerate plateaus are observed at $3N \times e^2/h$ values. $B - n$ phase space (Landau Fan) of $dG_{xx}/dn$ are shown at different values of $D/\varepsilon_0$. $3N \times e^2/h$ plateaus are marked with white arrows. Landau level crossings (red arrows) monotonically move towards higher $|n|$ values as $D/\varepsilon_0$ increases.

Here we identify Lifshitz transition in BLG by observing changes in the degeneracy of the low-energy Landau level spectrum. By measuring NLER we demonstrate that it changes sign near the charge neutrality point (CNP) and close to Lifshitz transitions, thereby detecting changes in Fermi surface topology. Our findings establish NLER as a powerful tool to track low energy Lifshitz transitions in materials with broken inversion symmetry, even at elevated temperatures ($T \geq 10$K) under time-reversal symmetric conditions. Our temperature-dependent characterization reveals that near the band edges, NLER is governed by an interplay between intrinsic and extrinsic mechanisms. The second-order nonlinear conductivity measured in our experiments exceeds 30 $\mu$mV$^{-1}\Omega^{-1}$, surpassing any previous reported values [6,7]. Our work provides new insights into the origin of NLER in BLG and establishes BLG as a highly tunable platform for generating NLER and exploring the physics of Lifshitz transitions.

The dual-gated hBN/BLG/hBN field effect transistors (FETs) were fabricated using the dry transfer technique [35,36] (See Supporting Information (SI) section 1 for fabrication and sample details). Fig. 1a depicts the schematic of the heterostructure. The measured resistivity

($\rho_{xx}$) at $T = 2$K is plotted as a function of carrier density ($n$) and vertical displacement field ($D$) in Fig. 1b (from sample Sa). We independently tune $n$ and $D$ by simultaneously controlling the top ($V_{tg}$) and bottom gate ($V_{bg}$) voltages (see SI section 1 for details). A monotonic increase of $\rho_{xx}$ with increasing $|D|$ indicates the band gap ($E_g$) opening due to inversion symmetry breaking [37]. Fig. 1c shows $\rho_{xx}$ vs $|n|$ data, on the hole side, at $D/\varepsilon_0 = -0.4$ Vnm$^{-1}$, confirming a low charge disorder density $\sim 10^{10}$ cm$^{-2}$ (vertical dashed line). The top left panel of Fig. 1d schematically shows the $E -$ momentum ($k$) dispersion of low energy conduction (CB) and valance bands (VB) of gapped BLG [27]. A projection of the CB on the vertical plane is shown in the top middle panel. The corresponding density of states (DoS) is displayed in the top right panel. Horizontal plane planes (i, ii, and iii) indicate cuts of the conduction band at three different fixed Fermi energies ($E_F$). The blue traces indicate the Fermi surfaces. Three low-energy Fermi pockets (in iii) merge into one (in i) via a Lifshitz transition at $E_F = \epsilon_L$, about $\sim 2$ meV away from the band edge [27]. The flat band at $\epsilon_L$ also leads to van Hove singularity (vHs) in the DoS. Furthermore, near $\epsilon_L$, the slope of the $E - k$ dispersion is reversed between the inner and outer Fermi surfaces, leading to opposite velocities and effective masses for the carriers on these surfaces, thereby resulting in carrier types with opposite polarities. The objective of this work is to investigate whether NLER is sensitive to these changes resulting from the Lifshitz transitions.

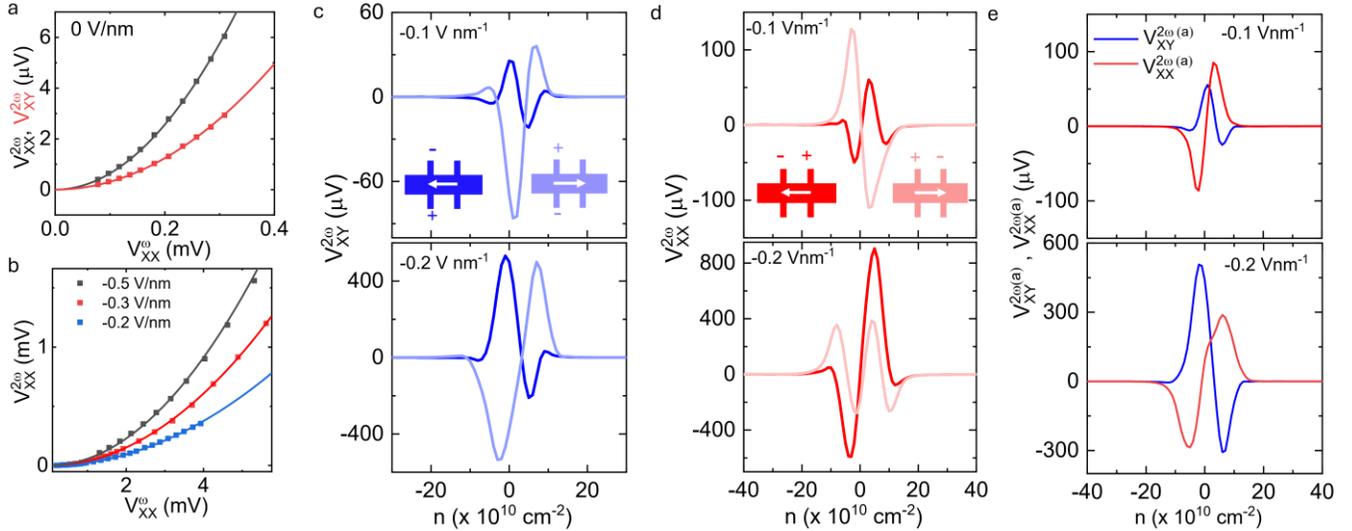

Figure 2. **Second order nonlinear electrical response (NLER) in BLG: a** and **b**, The parabolic dependence $V_{xy}^{2\omega}$ on $V_{xx}^{\omega}$ are presented at 0 and higher values of $D/\epsilon_0$, respectively. **c**, $n$ dependence of $V_{xy}^{2\omega}$ at $T = 10$ K with $I_{ds}^{\omega} = 100$ nA at two different values of $D/\epsilon_0$. The dark and light blue traces are data measured with opposite polarities of current and voltage probes, indicated schematically. **d**, simultaneously acquired $n$ dependent $V_{xx}^{2\omega}$ data are presented. Anti-symmetrized (with respect to measurement configurations) $V_{xy(x)}^{2\omega}$ vs $n$ data is shown as blue (red) traces at two different values of $D/\epsilon_0$.

We experimentally probe the Lifshitz transition in BLG by observing changes in the degeneracy of the Landau levels. Two probe conductance ($G_{xx}$) vs $n$ at $D/\epsilon_0 = 0.45$ Vnm$^{-1}$ under different vertical magnetic fields ($B$) from is shown in Fig. 1e (sample Sa). At higher values of $B$, $G_{xx}$ demonstrates plateaus at $N \times e^2/h$ values ($N$, $e$, and $h$ are integer, electronic charge, and Planck constant, respectively), corresponding to fully degeneracy-broken Landau levels. At lower values of B, we observe plateaus at $3N \times e^2/h$, corresponding to three Fermi pockets at $E_F < \epsilon_L$ [27,28]. Such changes in degeneracy, along with the emergence of a central Fermi island with opposite polarity, lead to a crossover of Landau levels [27,29] in the fan diagram of $dG_{xx}/dn$, far away from the disorders density as shown in Fig. 1f. The plateaus corresponding $3e^2/h$ and $6e^2/h$ conductance are marked with white arrows. Landau level crossings are marked with red arrows. These crossings are observed to shift monotonically toward higher $|n|$ values as $D/\epsilon_0$ increases, since $\epsilon_L$ rises with $D$.

After detecting the Lifshitz transition using first order transport we focus on NLER, which is inevitable in an inversion symmetry-broken system [5,10]. Second-order current density is given by $j_a^{2\omega} = \sigma_{abc}^{2\omega} E_b^{\omega} E_c^{\omega}$. Here $a, b, c$ are indices for axes. $E^{\omega}$ is the applied first-order bias. The $\sigma_{abc}^{2\omega}$, third rank tensor for second-order conductivity, is the property of the system and independent of its dimensions or applied bias. Experimentally, we apply a constant source-drain current bias ($I_{ds}^{\omega}$) of $\omega$ frequency and simultaneously detect transverse ($V_{xy}^{\omega}, V_{xy}^{2\omega}$) and longitudinal ($V_{xx}^{\omega}, V_{xx}^{2\omega}$) voltage drops at $\omega$ and $2\omega$ frequencies using lock-in technique. The black (red) data points in Fig. 2a represent $V_{xx}^{\omega}$ vs $V_{xx(y)}^{2\omega}$ data at different $I_{ds}^{\omega}$ (sample Sd), at $D = 0$ and $n = 2 \times 10^{10}$ cm$^{-2}$(also see SI section 2). Black, red and blue points in Fig. 2b shows $V_{xx}^{\omega}$ vs $V_{xx}^{2\omega}$ data at higher values of $D/\epsilon_0$ at $n = -25, -8$, and $-6 \times 10^{10}$ cm$^{-2}$, respectively (sample Sa). The solid lines in Fig. 2a and b are parabolic fits to $V_{xx(y)}^{2\omega}$. The $n$ dependence of $V_{xy}^{2\omega}$ at $D/\epsilon_0 = -0.1$ and $-0.2$ Vnm$^{-1}$ are demonstrated in Fig. 2c (measured with $I_{ds}^{\omega} = 100$ nA at $T = 10$ K, from sample Sb). The dark and light blue traces having opposite signs, are acquired in measurement configurations $\theta$ (left inset) and $\theta + \pi$ (right inset), with opposite current and voltage probe polarities. The NLER has two distinctive features, (i) $V_{xy(x)}^{2\omega}$ is proportional to $V_{xx}^{\omega 2}$, and (ii) $V_{xy(x)}^{2\omega}$ changes sign if the polarities of the voltage and current probes are switched simultaneously [6–13,38]. The agreement of data shown in Fig. 2a, b and c with the distinctive features (i) and (ii) unambiguously proves NLER as the origin of the observed $2\omega$ voltage response. Simultaneously acquired $V_{xx}^{2\omega}$ vs $n$ is presented in Fig. 2c (See SI section 3 for n dependence at larger $D$ from sample Sa). Notably, $V_{xy(x)}^{2\omega}$ contains a $n$ dependent component which does not change sign upon switching configurations from $\theta$ to $\theta + \pi$. This can originate from disorder hindering current flow in straight line, or due to thermal effects. We remove such effects by anti-symmetrizing and use $V_{xy(x)}^{2\omega(a)} = (V_{xy(x)}^{2\omega}|_\theta - V_{xy(x)}^{2\omega}|_{\theta+\pi})/2$ for future discussions. Fig. 2d demonstrates the $V_{xy(x)}^{2\omega(a)}$ vs $n$ data as blue(red) traces. Interestingly, $V_{xy(x)}^{2\omega(a)}$ changes sign at $n = 0$ detecting the transition of $E_F$ between valence and conduction band , similar to previous reports in other graphene based systems [7,38] .

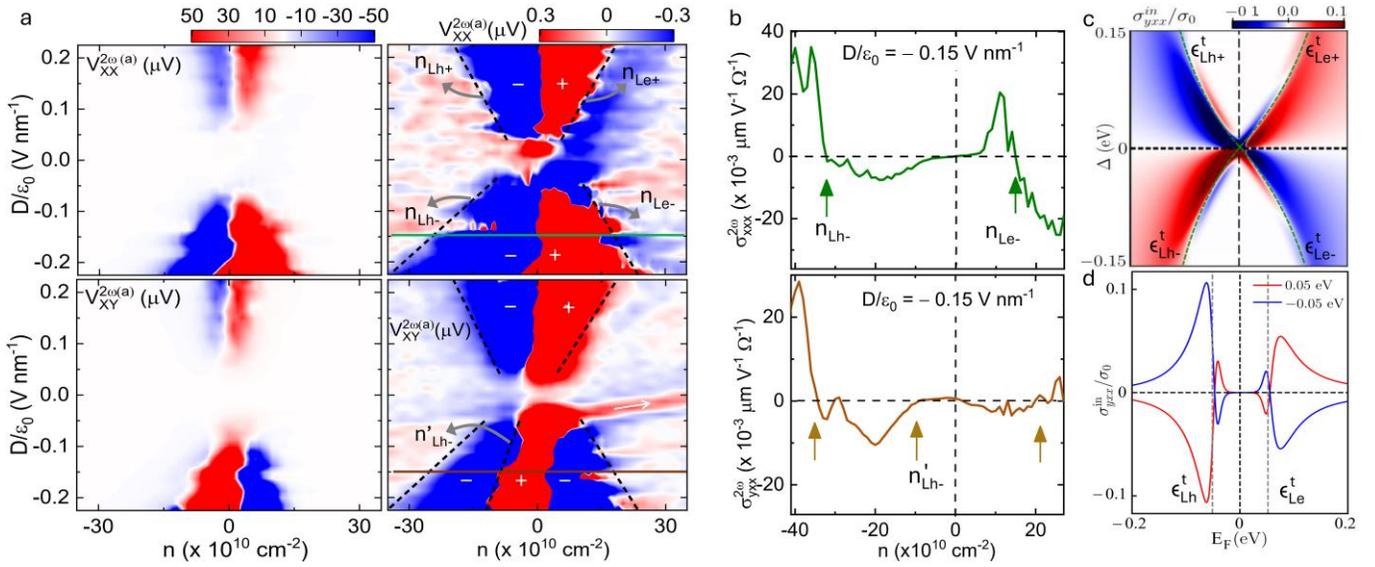

Figure 3. **Evidence of Lifshitz transition in NLER of BLG: a,** $n - D/\epsilon_0$ phase spaces of $V_{xx(y)}^{2\omega(a)}$ at $T = 10$ K with $I_{ds}^\omega = 100$ nA is shown in the top (bottom) left panel. The top (bottom) right panel presents the data with a reduced color scale. The black dashed lines ($n_{Lh\pm}$and $n_{Le\pm}$) indicate sign changes in NLER capturing in the vicinity of the Lifshitz transition. $n'_{Lh-}$ in the bottom right panel indicates an additional change in $V_{xy}^{2\omega(a)}$. Close to CNP ($|n| \leq 15 \times 10^{10}$cm$^{-2}$), the $D$ dependence of $V_{xy}^{2\omega(a)}$ is distinctly different from that of $V_{xx}^{2\omega(a)}$, indicating additional intrinsic contribution in $V_{xy}^{2\omega(a)}$. **b,** The top (bottom) panel shows $\sigma_{xxx}^{2\omega}$ ($\sigma_{yxx}^{2\omega}$), along the green (brown) horizontal trace of right top (bottom) panel of **a**. The sign changes in $\sigma^{2\omega}$ are indicated with arrows **c,** Calculated $\Delta - E_F$ phase space of intrinsic $2\omega$ conductivity $\sigma_{yxx}^{in}$ for a uniaxially strained BLG. E corresponding to Lifshitz transitions ($\epsilon_{h\pm}^t$ and $\epsilon_{e\pm}^t$) are marked by green dashed lines. **d,** Calculated $\sigma_{yxx}^{in}$ vs $E_F$ at $\Delta = \pm 0.05$ eV is presented, demonstrating sign changes at the Lifshitz transition.

To understand the evolution of NLER with band gap opening, and possible role of Lifshitz transition, we now focus on $n$ vs $D/\epsilon_0$ phase spaces of $V_{xx}^{2\omega(a)}$ and $V_{xy}^{2\omega(a)}$ from the same sample (Sb), which are presented in the top and bottom left panels of Fig. 3a, respectively (at $I_{ds}^\omega = 100$ nA and $T = 10$ K). See SI section 4 for $V_{xy(x)}^\omega$ data. Intriguingly, at positive (negative) $D$, the sign of $V_{xy}^{2\omega(a)}$ is observed to switch from negative (positive) to positive (negative) as $n$ is increased from negative to positive close to CNP. Clearly in this regime ($|n| \leq 15 \times 10^{10}$cm$^{-2}$), unlike $V_{xx}^{2\omega(a)}$, the sign of $V_{xy}^{2\omega(a)}$ is dependent on the sign of $D$. Such distinct $D$ dependence points towards a difference in the origin of $V_{xy}^{2\omega(a)}$, the transverse NLER. To further elucidate the salient features in NLER, we present the data with reduced levels of the color scale in the top and bottom right panels of Fig. 3a. In addition to the sign change close to CNP, marked as $+$ and $-$ lobes, $V_{xx}^{2\omega(a)}$ also demonstrate sign changes across black dashed lines marked as $n_{Le\pm}$ and $n_{Lh\pm}$, before vanishing into experimental noise. Here, the indices $e(h)$ and $+(-)$ refer to electron (hole) and positive (negative) $D$ regime. In $V_{xy}^{2\omega(a)}$, we observe an emergent sign change at $n'_{Lh-}$. This indicates an additional emergent intrinsic NLER having $+ - -$ and $- - +$ lobes at negative and positive $D$, respectively. The feature marked with a white arrow arises from the non-top gated BLG contacts. In the top (bottom) panel of Fig. 3b, we show calculated second-order longitudinal (transverse) conductivity $\sigma_{xxx}^{2\omega}(\sigma_{yxx}^{2\omega}) = \frac{\sigma V_{xx}^{2\omega(a)}L}{V_{xx}^{\omega 2}}$ ($\frac{\sigma V_{xy}^{2\omega(a)}L^2}{W V_{xx}^{\omega 2}}$) along the green (brown) horizontal sections in top (bottom) right panel of Fig. 3a. Here $L$, $W$ and $\sigma$ are the length, and width of the channel and first order conductivity, respectively. The sign changes in $\sigma^{2\omega}$ are indicated with arrows. Notably, the $\sigma^{2\omega}$ is minuscule close to CNP, but increases sharply across $n_{Le\pm}$ and $n_{Lh\pm}$. The $\sigma^{2\omega}$ is related to the Fermi distribution function $f(k)$ through $\frac{\partial^2 f(k)}{\partial k_a \partial k_b}$ and $\frac{\partial f(k)}{\partial k_a}$ [5], which alters sign and magnitude close to Lifshitz transitions in BLG, particularly at the inner Fermi island with opposite carrier polarity. Our theoretically calculated skew scattering contribution to $\sigma_{xxx}^{2\omega}$ (see SI section 11) demonstrates sign changes at only at CNP and in the vicinity of Lifshitz transitions. Furthermore, experimentally observed values of $n_{Le\pm}$ and $n_{Lh\pm}$ closely matches with previous reports in BLG at similar interlayer asymmetry [39]. Our experimental and theoretical study confirms that $n_{Le\pm}$ and $n_{Lh\pm}$ are related to Lifshitz transition (Also see SI section 5) for data on sign change in NLER away from CNP from sample Sa). Thus, NLER in BLG can probe the low-energy Lifshitz transitions even at elevated $T \geq 10$ K.

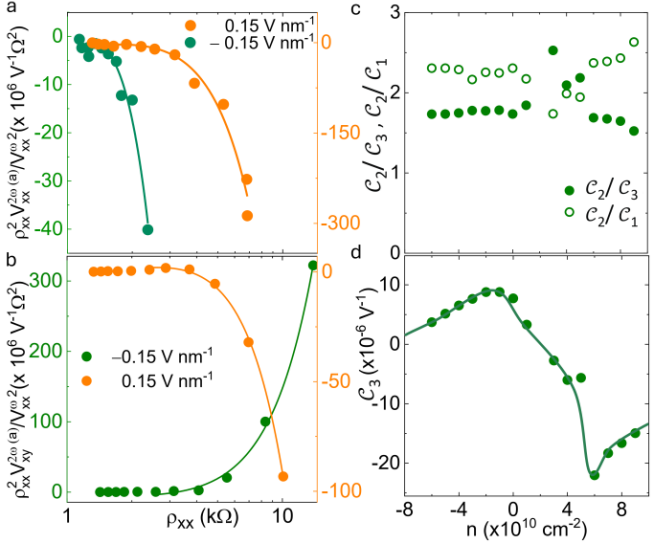

Figure 4. **Origin of NLER in BLG from scaling theory: a,** $\rho_{xx}^2 V_{xx}^{2\omega}/V_{xx}^{\omega 2}$ vs $\rho_{xx}$ data, two different $D/\varepsilon_0$ values. The solid lines are the fits using the generalized scaling equation, indicating an interplay of extrinsic side jump and skew scattering mechanisms. **b,** $\rho_{xx}^2 V_{xy}^{2\omega}/V_{xx}^{\omega 2}$ vs $\rho_{xx}$ data and fits at two different $D/\varepsilon_0$ values are shown. **c,** $\mathcal{C}_2/\mathcal{C}_3$ and $\mathcal{C}_2/\mathcal{C}_1$ from fits of $\rho_{xx}^2 V_{xy}^{2\omega}/V_{xx}^{\omega 2}$ vs $\rho_{xx}$ data at $D/\varepsilon_0 = -0.15$ Vnm$^{-1}$. In the scaling theory, the ratio of weightage of intrinsic NLER in the scaling parameters $\mathcal{C}_1$, $\mathcal{C}_2$, and $\mathcal{C}_3$ is 1:2:1. Observed $\mathcal{C}_2/\mathcal{C}_3 \approx \mathcal{C}_2/\mathcal{C}_1 \approx 2$ indicates a dominant intrinsic contribution in $V_{xy}^{2\omega}$. **d,** $\mathcal{C}_3$ vs $n$ data at $D/\varepsilon_0 = -0.15$ Vnm$^{-1}$ is presented, demonstrating the sign change of intrinsic component of $V_{xy}^{2\omega}$ at CNP. The solid line is a guide to the eye.

The $D$ switchable intrinsic $V_{xy}^{2\omega(a)}$ arising from BCD [12,38] can be enabled by the uncontrolled interfacial heterostrain appearing during fabrication of vdW heterostructures [40–46]. Raman spectra from the measured samples (Sb and Sc) confirms the presence of up to 0.6% heterostrain (se SI section 7). We theoretically calculate the intrinsic $2\omega$ conductivity $\sigma_{yxx}^{in}$ of BLG under 1% uniaxial strain (see SI section 11). Fig. 3c present $E_F - \Delta$ phase space of $\sigma_{yxx}^{in}$, where $\Delta \approx e d_{vdw} D/\varepsilon_0$ is the interlayer potential, $e$ is the electronic charge and $d_{vdw} \approx 0.3$ nm. The green dashed traces denote the evolution of Lifshitz transition with increasing $|\Delta|$. $\sigma_{yxx}^{in}$ demonstrates clear sign change close to the Lifshitz transitions and CNP. Additionally, $\sigma_{yxx}^{in}$ also switches sign when the sign of $\Delta$ (or $D$) is changed, matching experimental $V_{xy}^{2\omega}$. Fig. 3d represents $\sigma_{yxx}^{in}$ vs $E_F$ data at $\Delta = \pm 0.05$ eV. Here, the grey dashed vertical lines indicate the locations of Lifshitz transitions. Our calculations qualitatively agree with experimentally observed $V_{xy}^{2\omega(a)}$ close to CNP.

This indicates a significant additional intrinsic contribution to $V_{xy}^{2\omega(a)}$ close to CNP. To understand the sample-to-sample variations of our observations, we measured $n$ vs $D/\varepsilon_0$ phase space of $V_{xx}^{2\omega(a)}$ and $V_{xy}^{2\omega(a)}$ from a different sample (sample Sc ) hosting uniaxial heterostrain (see Raman data in SI section 7). The data is presented in SI section 6, which qualitatively match with the data from sample Sb.

To further elucidate the contributions of intrinsic BCD and extrinsic mechanisms in NLER, we now focus on its $\rho_{xx}$ dependent scaling behavior. We use $T$ as a tuning parameter for $\rho_{xx}$ keeping $D$ and $n$ values fixed. SI section 8 shows $V_{xx}^{\omega}$ and $V_{xy(x)}^{2\omega(a)}$ vs $n$ data at different $T$. At first, we focus on $V_{xx}^{2\omega(a)}$ which only contains extrinsic contribution. The green and the orange points in Fig. 4a illustrate $\rho_{xx}^2 V_{xx}^{2\omega(a)}/V_{xx}^{\omega 2}$ vs $\rho_{xx}$ data for $D/\varepsilon_0 = -0.15$ Vnm$^{-1}$ and $0.15$ Vnm$^{-1}$ at $n \approx -10 \times 10^{10}$ cm$^{-2}$, respectively. The data is well-fitted (green and orange solid lines) by the generalized disorder-induced scaling behavior of NLER (see SI section 10 for more details), $\rho_{xx}^2 \frac{V_{xy(x)}^{2\omega(a)}}{V_{xx}^{\omega 2}} = \mathcal{C}_1 \rho_0^2 + \mathcal{C}_2 \rho_0 \rho_T + \mathcal{C}_3 \rho_T^2$. Here, $\rho_0 = \rho_{xx}(T \to 0)$, and $\rho_T = \rho_{xx}(T) - \rho_0$. $\mathcal{C}_1$, $\mathcal{C}_2$, and $\mathcal{C}_3$ are scaling parameters containing contributions of intrinsic BCD (appearing only in $V_{xy}^{2\omega}$), side jump and skew scattering from static and dynamic impurities. The intrinsic contribution has twice the weightage in $\mathcal{C}_2$, compared to that in $\mathcal{C}_1$ and $\mathcal{C}_3$ [5]. For orange (green) fits (in Fig. 4a), values of $\mathcal{C}_1$, $\mathcal{C}_2$, $\mathcal{C}_3$ are $-5.5 \times 10^{-6} \pm 2.4 \times 10^{-7}$ ($-6.9 \times 10^{-6} \pm 3.9 \times 10^{-7}$) V$^{-1}$, and $-9.6 \times 10^{-6} \pm 1.8 \times 10^{-6}$ ($-34.2 \times 10^{-6} \pm 5.1 \times 10^{-6}$) V$^{-1}$, respectively. Clearly, each parameter has the same orders of magnitude, indicating the presence of the side jump mechanism in addition to skew scattering from static and dynamic impurities. Likewise, Fig. 4b presents $\rho_{xx}^2 V_{xy}^{2\omega}/V_{xx}^{\omega 2}$ vs $\rho_{xx}$ at positive (negative) $D$ as orange (green) data points at $n = -0.03 \times 10^{12}$ cm$^{-2}$. In this case the values of $\mathcal{C}_1$, $\mathcal{C}_2$, $\mathcal{C}_3$ are obtained to be $-1.5 \times 10^{-6} \pm 0.8 \times 10^{-7}$ ($1.6 \times 10^{-6} \pm 3.6 \times 10^{-8}$) V$^{-1}$, $-3.2 \times 10^{-6} \pm 1.9 \times 10^{-7}$ ($3.8 \times 10^{-6} \pm 1.1 \times 10^{-7}$) V$^{-1}$ and $-1.7 \times 10^{-6} \pm 1.1 \times 10^{-7}$ ($2.2 \times 10^{-6} \pm 2.4 \times 10^{-7}$) V$^{-1}$, respectively. Interestingly, here the ratio $\mathcal{C}_1 : \mathcal{C}_2 : \mathcal{C}_3$ is observed to be $\approx 1 : 2 : 1$, for both fits. This indicates a significant contribution from the intrinsic BCD in the NLER in our BLG heterostructure close to CNP. Fig. 4c presents $\mathcal{C}_2/\mathcal{C}_3$ and $\mathcal{C}_2/\mathcal{C}_1$ as a function of $n$. Fig. 4d presents $\mathcal{C}_3$ vs $n$ demonstrating a sign change at CNP, consistent with our theoretical calculations for intrinsic contribution. Additionally, in the SI section 12, we show NLER data at higher $n$, measured with higher $I_{ds}^{\omega}$ to reduce experimental noise (from sample sB). We obtain $\sigma_{xxx}^{2\omega} > 30 \; \mu$mV$^{-1}\Omega^{-1}$ at $n = -70 \times 10^{10}$ cm$^{-2}$, at $D/\varepsilon_0 = -0.125$ Vnm$^{-1}$ at $T = 3$K which is higher than the values reported for any other 2D vdW materials (see SI section 9 ) or their heterostructure to the best of our knowledge [6,7].

In conclusion, combining experiment and theoretical calculations, we demonstrated that NLER detects the sudden change in the Fermi surface topology at the Lifshitz transition near the band edge of BLG by undergoing a sign reversal. Our study reveals that NLER can be used as a general tool to locate Lifshitz transitions in inversion symmetry broken materials, even at elevated temperatures. Moreover, we show that the intrinsic Berry curvature dipole arising from interfacial strain dominates the transverse NLER close to the charge neutrality point (CNP). Furthermore, the measured second harmonic conductivity in BLG exceeds 30 $\mu m V^{-1}\Omega^{-1}$, higher than that of any other 2D van Der Waal materials or their heterostructures at 2K. Our work establishes BLG as a highly tunable platform for NLER and its applications.

**Associated content:**

**Data availability statement:** All data needed to evaluate to the conclusions in the paper are present in the paper and/or the supporting information. All source data related to the findings in this study can be obtained from the authors. All the experimental data will be uploaded to the Zenodo repository upon publication.

**Supporting information:** The Supporting information (SI) is available. Supporting information contains the following sections. Section 1: Fabrication and device details and measurement methods. Section 2: $I_{ds}^{\omega}$ dependent $V_{xy}^{2\omega}$ vs $n$ data (sample Sd) Section 3: $V_{xx}^{2\omega}$ vs $n$ data at larger $D$ (sample Sa). Section 4: $D/\epsilon_0 - n$ phase space of $V_{xy(x)}^{\omega}$ (sample Sb). Section 5: $D/\epsilon_0 - n$ phase space of $V_{xy}^{2\omega}$ (sample Sa). Section 6: $D/\epsilon_0 - n$ phase spaces of $V_{xx}^{\omega}$, $V_{xx}^{2\omega(a)}$ and $V_{xy}^{2\omega(a)}$ (sample Sc). Section 7: Evidence of strain from sample Sb and Sc. Section 8: $V_{xy}^{2\omega(a)}$ vs $n$ data at different $T$ from sample Sb. Section 9: Estimation of $\sigma_{xxx}^{2\omega}$ and $\sigma_{yxx}^{2\omega}$ (sample Sc) Section 10: Scaling theory of the nonlinear electrical currents. Section 11: Calculation of Band structure and nonlinear conductivity in strained bilayer graphene.

**Author contributions:** T.A. and L.E.H. conceived the project. T.A. fabricated the samples, performed the measurements, and analyzed the data with assistance from B.Q.T. H.V. and A.A. performed the theory calculations. T.T. and K.W. provided the hBN crystals. T.A. and H.V. wrote the experiment and theory parts of the paper. All authors discussed, commented on the paper. **note:** The authors declare that they have no competing interests. **Acknowledgements:**
The authors acknowledge financial support from MICIU/AEI/10.13039/501100011033 (Grant CEX2020-001038-M), from MICIU/AEI and ERDF/EU (Projects PID2021-122511OB-I00 and PID2021-128004NB-C21), from MICIU/AEI and European Union NextGenerationEU/PRTR (Grant PCI2021-122038-2A) and from the European Union (Project 101046231-FantastiCOF). T.A. acknowledges funding from the European Union under Marie Sklodowska-Curie grant agreement number 101107842 (ACCESS). H.V. acknowledges the Ministry of Education, Government of India, for funding support through the Prime Minister's Research Fellowship program. K.W. and T.T. acknowledge support from the JSPS KAKENHI (Grant Numbers 21H05233 and 23H02052) and World Premier International Research Center Initiative (WPI), MEXT, Japan.

* t.ahmed@nanogune.eu
† amitag@iitk.ac.in
‡ l.hueso@nanogune.eu